\documentclass[floatfix,amsmath,amsfonts]{revtex4-1}
\usepackage[latin9]{inputenc}
\usepackage{color}
\usepackage{amsmath}
\usepackage{graphicx}

\makeatletter
\@ifundefined{textcolor}{}
{%
 \definecolor{BLACK}{gray}{0}
 \definecolor{WHITE}{gray}{1}
 \definecolor{RED}{rgb}{1,0,0}
 \definecolor{GREEN}{rgb}{0,1,0}
 \definecolor{BLUE}{rgb}{0,0,1}
 \definecolor{CYAN}{cmyk}{1,0,0,0}
 \definecolor{MAGENTA}{cmyk}{0,1,0,0}
 \definecolor{YELLOW}{cmyk}{0,0,1,0}
}

\usepackage{color}
\usepackage{graphics}
\usepackage{epsfig}
\newcommand{\be}{\begin{equation}}
\newcommand{\ee}{\end{equation}}
\newcommand{\bea}{\begin{eqnarray}}
\newcommand{\eea}{\end{eqnarray}}
\newcommand{\bes}{\begin{subequations}}
\newcommand{\ees}{\end{subequations}}

\makeatother

\begin{document}

\title{Bloch oscillations sustained by nonlinearity}

\author{R. Driben$^{1}$, V. V. Konotop$^{2}$, T. Meier$^{1}$, and A. V. Yulin$^{3}$}
\affiliation{
$^{1}$Department of Physics and CeOPP, University of Paderborn, Warburger Str. 100, D-33098 Paderborn, Germany\\
 $^{2}$ Centro de F\'{i}sica Te\' orica e Computacional  and Departamento de F\'{i}sica, Faculdade de Ci\^encias, Universidade
de Lisboa, Campo Grande, Ed. C8, Lisboa 1749-016, Portugal\\
 $^{3}$ITMO University, 49 Kronverskii Ave., St. Petersburg 197101, Russian Federation\\Correspondence and requests for materials should be addressed to R.D. (email:driben@mail.uni-paderborn.de)}

\date{\today}

\begin{abstract}
 We demonstrate that nonlinearity may play a constructive role in supporting Bloch oscillations in a model which is discrete, in one dimension and continuous in the orthogonal one. The model can be experimentally realized in several fields of physics such as optics and Bose-Einstein condensates.
We demonstrate that designing an optimal relation between the nonlinearity and the linear gradient strength provides extremely long-lived Bloch oscillations with little degradation. Such robust oscillations can be observed for a broad range of parameters and even for moderate nonlinearities and large enough values of linear potential.
We also present an approximate analytical description of the wave packet's evolution featuring a hybrid Bloch oscillating wave-soliton behavior that excellently
corresponds to the direct numerical simulations.
\end{abstract}

\flushbottom
\maketitle
%
%
\thispagestyle{empty}


\maketitle
\section*{Introduction}
The celebrated phenomenon of Bloch oscillations~\cite{Zener, Zener2} (BOs) was originally proposed for electrons in crystals
in the presence of homogeneous electric fields which give rise to a potential that varies linearly in the field direction.
After a long lasting debate about the actual existence of BOs, see, e.g., \cite{rabin, rabin2}, rigorous upper bounds for the
interband tunelling rates could be established and the effective Hamiltonians that lead to BOs and their frequency-domain
counterpart the Wannier-Stark ladder could be justified, see, e.g., \cite{nenciu} and references therein.
In the early 1990s BOs were first observed experimentally in electrically-biased semiconductor superlattices
using optical interband excitation with femtosecond laser pulses~\cite{jochen, jochen2}.
A few years later, also for atoms in optical lattices~\cite{atoms} and for coupled waveguides~\cite{peschel}
BOs have been realized.
This proves that BOs can be considered in a broader context as a fundamental effect that may occur in systems which
support wave propagation in media with periodically-varying parameters and with a linear potential.

The physical understanding of BOs comes from the band-gap structure of the underlying periodic linear potential
and can be viewed as a Bloch mode "motion" along the dispersion curve~\cite{Zener, Houston}.
In addition to the existence of the band-gap structure such an interpretation requires the linear gradient to be small
(otherwise it cannot be accounted in terms of the adiabatic theorem and must be considered in leading order). Thus by
its nature BO is a linear phenomenon and it is common belief that nonlinearity plays a destructive role
which makes it impossible to observe BOs at long times (or propagation distances, depending on the particular physical context)
even without dephasing processes.
This  was first reported in~\cite{BishopSalerno} and later on confirmed experimentally in optics using arrays of Kerr-type waveguides~\cite{Nonlin_optics} and furthermore in Bose-Einstein condensates (BECs) loaded in optical lattices~\cite{Nonlin_BEC_1, Nonlin_BEC_1a, Nonlin_BEC_2}, where only a few oscillations were detected. The main reason which suppresses long-living nonlinear BOs was discussed in~\cite{BKS} and originates from the modulation instability of Bloch waves at different edges of the band gap, where the effective mass (effective dispersion) changes its sign: Bloch waves are  stable and unstable at the opposite edges provided the nonlinearity remains constant~\cite{KS}.
This understanding has led to several suggestions of rather complicated spatial~\cite{spatial, spatial2} and temporal~\cite{temporal, temporal2} nonlinear management techniques which could support long-lived BOs. All of such proposals are based on the idea of changing the sign of the effective nonlinearity synchronized with the change of the sign of the effective mass in a way that their product remains of the same sign during the evolutions. This requires controlled modification of the system's properties.

Considering BOs as a broader concept, namely as the periodic evolution of systems obeying a discrete translational invariance and being subject to a linear gradient,
they exist even in strongly nonlinear systems and in the presence of an arbitrary large gradient, if the system is exactly integrable. This has been obtained analytically~\cite{Bruschi,KCV} and numerically~\cite{Bishop} for integrable discrete nonlinear Schr\"odinger equations (known also as the Ablwitz-Ladik model~\cite{AL}), as well as for its integrable generalizations~\cite{Vakhnenko}. While the mathematical reason for the exact periodic motion of such systems consists in their exact integrability, the physical explanation relies on the property of a specific nonlocal nonlinearity in such models which leads to stable Bloch modes at both band edges~\cite{BKS}.

 In the integrable models the phenomenon of BOs is not restricted to small amplitudes of the linear potential. When the potential strength becomes large enough the pulses become practically localized in space since the amplitude of BOs can become less than the width of the pulse. A similar non-spreading behavior of wave packets can be also observed in non-integrable models at large nonlinearities~\cite{Kolovsky}. On the other hand, when the strength of the nonlinearity increases, the non-integrable models show other types of behavior
 like the transient phenomenon of single-site trapping followed by explosive spreading and subdiffusion of the wave packet~\cite{Krimer}.

So far, no non-integrable systems with a \emph{constant} nonlinearity coefficient, which support long-living BOs, have been proposed.
Here, we fill this gap and introduce and analyze a physically-relevant non-integrable model which
does show BO dynamics persisting for long times at considerable nonlinearities and linear gradients.
As it is demonstrated below, balance between the effects of the nonlinearity and the dispersion can be achieved in systems that contain an additional dimension besides the dimension corresponding to the direction of the linear gradient.  This balance
may result in the existence of very stable oscillatory motion of discrete-continuous soliton-like wave packets.

\section*{RESULTS}

\subsection*{ Model and Linear Dynamics}

We consider an array of coupled one-dimensional nonlinear waveguides which are subject to a linear potential. Thus our system is effectively two-dimensional with one discrete and one continuous spatial variables. It is described by the coupled nonlinear Schr\"odinger equations which in dimensionless variables read
\begin{eqnarray}
 \label{eq:inf_chain}
 i\frac{\partial u_n}{\partial t}+\alpha\frac{\partial^2 u_n}{\partial x^2} + \kappa(u_{n-1}+u_{n+1}
 -2u_n)
 +\gamma nu_{n} +g|u_n|^2u_n=0 \, .
 \end{eqnarray}
Here $u_n(t,x)$ is the nonlinear field, $\kappa$ is the coupling between neighbour waveguides,  $\alpha$ is the continuous diffraction coefficient, $\gamma$ is the strength of the linear gradient, and $g$ is the nonlinearity which is considered to be attractive
(or focusing, depending on the physical context), i.e., $g\geq 0$.

Equation~\eqref{eq:inf_chain} describes the light propagation in an array of coupled optical fibers~\cite{opt_appl} in the presence of a linear gradient of the waveguide effective index.
In this case, $u_n$ is the dimensionless electric field, the evolution coordinate $t$ needs to be interpreted as the spatial coordinate along the fiber, and $x$ will be the normalized retarded time. Thus Eq.~\eqref{eq:inf_chain} properly describes the evolution of an optical pulse experiencing continuous dispersion together with discrete diffraction in presence of  Kerr nonlinearity and a linear gradient of the waveguide effective index.
The model defined by Eq.~(\ref{eq:inf_chain}) is even more generic.
In addition, it also describes an array of coupled quasi-one-dimensional BECs, where $u_n$ stands for the dimensionless order parameter in $n$-th trap minimum. In the experiment the respective traps can be created by deep periodic optical lattices, see, e.g., \cite{Bloch, Bloch2}. In such a statement the discrete index $n$ numbers the successive minima of the optical lattice and $\kappa$ characterizes coupling due to the tunneling of atoms between neighbor minima.  Such a model can be viewed as extensions of a previous study~\cite{Malomed_coupled} of two BEC array created by a double-well potential, to the case of a trap created by an optical lattice.

Since BOs were discovered and are usually considered to be purely linear phenomenon, where in one-dimensional settings the nonlinearity plays a destructive role, we start with the linear case and set $\alpha=0.5$ and $g=0$. In this limit the Cauchy problem defined by Eq.~\eqref{eq:inf_chain} supplied by the initial condition $u_m^{(0)}(x)=u_m(x,t=0)$, can readily be solved explicitly (with tilde we denote the linear limit):
 \begin{eqnarray}
 \label{eq:linear}
 \tilde{u}_n=
 \frac{1-i }{2\sqrt{\pi t}} \sum_m  (-1)^{n-m} e^{i(\gamma m-2\kappa  ) t}J_{n-m}
 \left(\frac{2\kappa}{\gamma}\right)
 \int_{-\infty}^{\infty}\exp\left[i\frac{(x-\xi)^2}{2t}\right]u_{m}^{(0)}(\xi)d\xi ,
 \end{eqnarray}
where $J_n(\cdot)$ is the $n$-th order Bessel function.

\begin{figure}
 	\includegraphics[width=12cm]{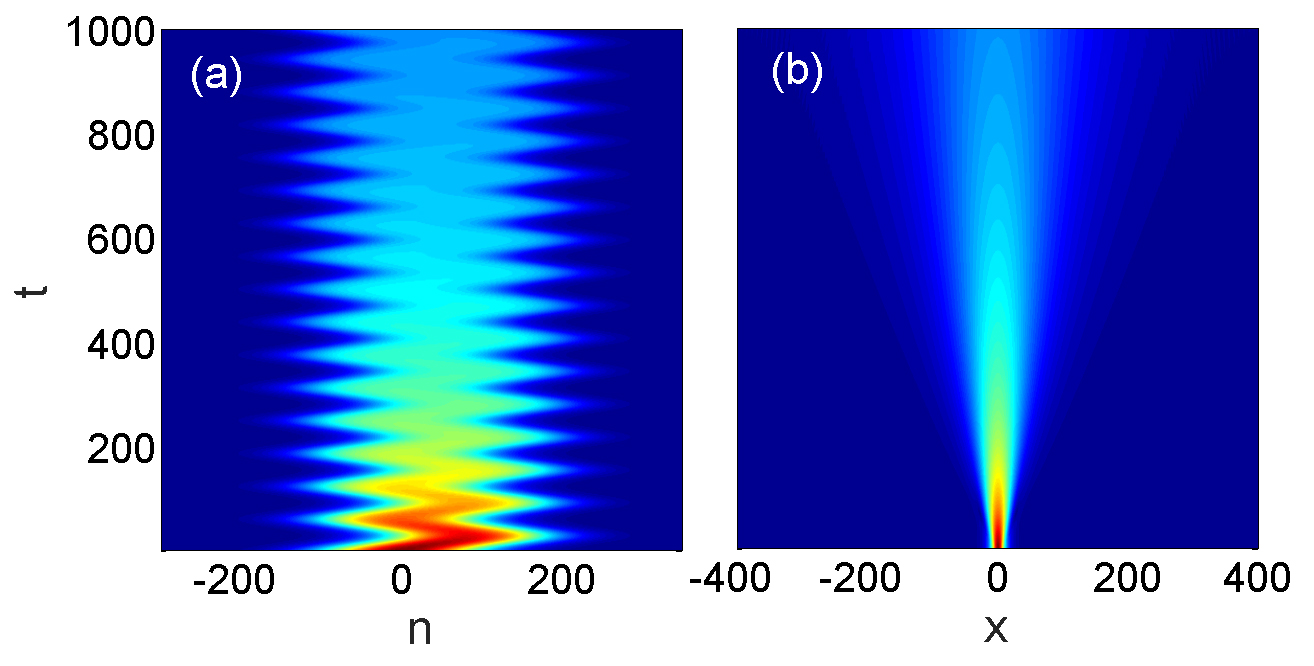}
		\caption{Propagation of a wave packet in a linear system, i.e., for $g=0$.
(a) and (b) show the evolution of the wave packet in the $n$-$t$ and in the $x$-$t$ planes, respectively.
The parameters are chosen as $\alpha=0.5$, $\kappa=2$ and $\gamma=0.1$.
The initial condition is given by Eq.~(\ref{init}) with $a_0=0.15$ and $w=100$.
The initial condition is chosen to be wide along $n$ to make this case close to typical BOs. Hereafter we display the modulus of the field- $|{\emph{u}}_n|$.
}
\label{fig:one}
\end{figure}

For the sake of definiteness, in all numerical simulations presented below  we consider initial conditions having a Gaussian envelope with respect to $n$ and sech-like profiles with respect to $x$:
 \begin{eqnarray}
 \label{init}
 u_n^{(0)}(x)=\frac{A\left(n\right)}{\cosh\left[A\left(n\right)x\right]}, \,\,\mbox{where}\,\,  A(n)=a_0e^{-n^2/w^2},
 \end{eqnarray}
 $w$ is the characteristic width of the initial wave packet along $n$-direction, and $a_0$ characterizes the wave packet amplitude.
In Fig.~\ref{fig:one} we illustrate the dynamical evolution of the linear solution according to Eqs.~(\ref{eq:linear}) and (\ref{init}).
	Panel (a) illustrates the oscillations of the wave envelope along the discrete coordinate with the amplitude and the frequency given by the analytic formulas. The significant decrease of the intensity of the field is clearly seen and is explained by the spreading of the envelope along $x$ coordinate [see Fig.~\ref{fig:one}(b)].

\begin{figure}
	\includegraphics[width=16cm]{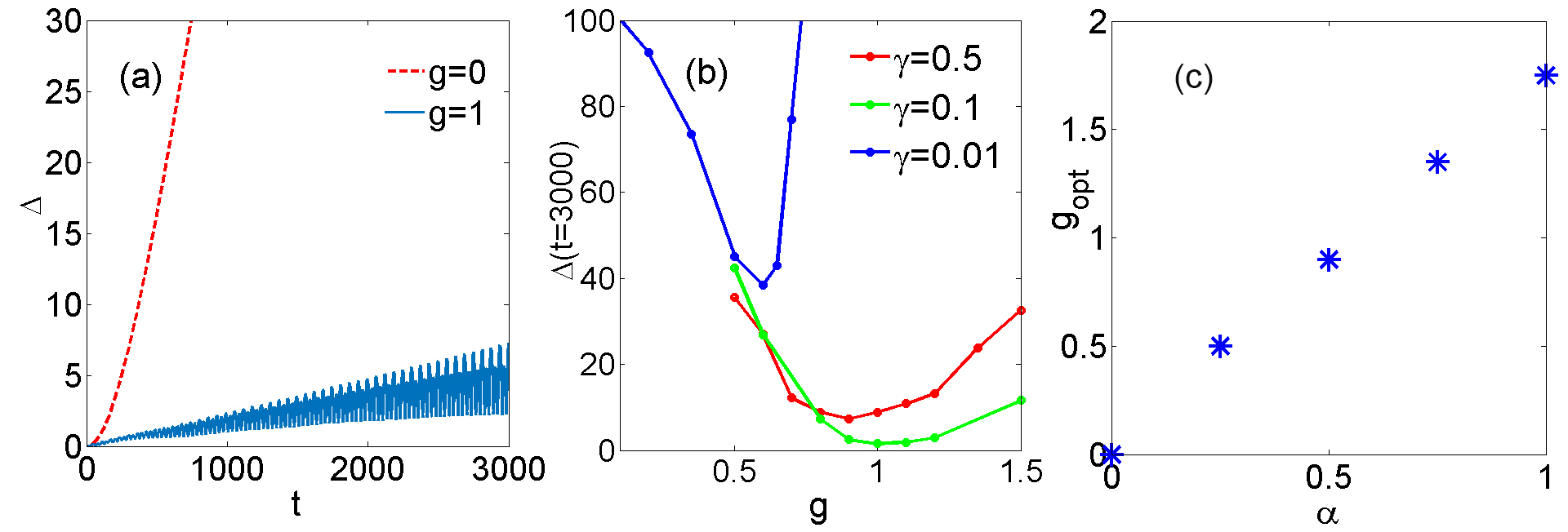}
	\caption{(a) shows the temporal evolution of the overall spread of the wave packet $\Delta(t)$ for the linear case pertaining to Fig.~\ref{fig:one} and to the nonlinear case shown in Fig.~3(a,b). (b) $\Delta(t=3000)$, i.e., the spread after a long time evolution, as function of the nonlinearity $g$ for different gradient coefficients: $\gamma=0.01$, $0.1$, and $0.5$. For (a) and (b) we set $\alpha=0.5$. (c) Optimal values of the nonlinear coefficient $g$ for various values of the continuous diffraction coefficient $\alpha$. The gradient strength is set to $\gamma=0.1$. }
	\label{fig:two}
\end{figure}

In order to characterize the dynamics of the wave-packet both in linear and (below) nonlinear cases we define the average of an arbitrary function $f_n(x,t)$ by the formula $\langle f \rangle=\frac{1}{P}\int_{-\infty}^{\infty}\sum_{n} f_n(x,t) |u_n(x,t)|^2dx$, where $P=\sum_n \int_{-\infty}^{\infty}|u_n(x,t)|^2 dx$. This allows us to explore the average positions of the wave along $x$ and $n$ directions, i.e., $\langle x \rangle$ and $\langle n \rangle$, respectively. Furthermore, we define the deformation parameter characterizing "combined" changes of the wave packet width of the wave packet during the evolution:
\begin{eqnarray}
\Delta(t)=\sqrt{[N(t)-N(0)]^2 + [X(t)-X(0)]^2} \, .
\end{eqnarray}
Here $N(t)=\sqrt{\langle (n-\langle n\rangle)^2\rangle}$ and $X(t)=\sqrt{\langle (x-\langle x\rangle)^2\rangle}$ are  the average widths of the wave packet in $n$ and $x$ directions.  {If deformations with respect to $n$ and $ x$ are strongly asymmetric, the parameter $\Delta$ is the estimate of the largest deformation of the wave envelope.}
For the ideal case of totally robust BOs, $\Delta(t)$ would be time independent. A growing  {or decreasing} deformation parameter $\Delta(t)$ corresponds to increasing deformations of the initial wave packet.
The introduced deformation parameter is shown in Fig.~\ref{fig:two} (a) and (b).
In particular, in full agreement with the evolution shown Fig.~\ref{fig:one}, the red dashed line in Fig.~\ref{fig:two} (a) illustrates the very rapid increase of $\Delta(t)$ in the linear case corresponding to the absence of long-lived BOs in this regime.

When a focusing nonlinearity is present, an obvious expectation is that it may compensate the diffraction leading to a slower spreading of the wave packet along the $x$-direction or eventually even to stationary propagation. Thus the nonlinearity would prevent the decay of the beam amplitude. On the other hand, one also expects the destruction of the BOs in the $n$-direction in the presence of a nonlinearity.
However, since the reasons for the decay of BOs in (weakly) nonlinear one-dimensional systems arise from the change of the effective diffraction (effective mass, using in solid state terminology)  when a beam moves between the two opposite edges of a band, one may expect that adding an additional direction may weaken this effect and thus stabilize BOs.

Indeed, the dispersion relation associated with the linear case of Eq.~(\ref{eq:inf_chain}) at $g=0$  is obtained by the ansatz $u_n(x,t)\sim e^{i(\omega t+qn+kx)}$ and reads $\omega=-k^2/2+4\kappa\sin^2(q/2)$. Thus near the center and the boundary of the Brillouin zone, i.e., at $|q|\ll 1$ and $q=\pi+\tilde{q}$ with $|\tilde{q}|\ll1$, respectively, the dispersion relation is given by $\omega\approx  -k^2/2+\kappa q^2$ and by $\omega\approx -k^2/2+4\kappa-\kappa \tilde{q}^2$. So, at the boundary of the Brillouin zone for a focusing nonlinearity the wave packet will be compressed along both directions since both  curvatures are negative. In a  continuous homogeneous medium with the parabolic dispersion relation $-k^2/2-\kappa q^2$  and Kerr nonlinearity there exists only the unstable Townes soliton and hence the discreteness preventing the collapse plays a stabilizing role. On the other hand, at the center of the Brillouin zone the curvatures along $k$ and $q$ directions ($\partial_x^2 \omega$ and $\partial_q^2 \omega$, correspondingly)  are of opposite signs. The one associated with discrete variable is positive and results in an effective dispersion tending to destroy the localized wave packet. The amplitude of this dispersive wave packet, however, does not decay as fast as it would happen in the $x$-independent case, since now the compression of the wave packet along the $x$-direction may compensate the decay of the wave packet amplitude due to the dispersion. These simple qualitative arguments allow us to suggest that the interplay of the nonlinearity with the discreteness of the model Eq.~(\ref{eq:inf_chain}) may enhance the stability of nonlinear BOs allowing them to become long-lived.

\begin{figure}
	\includegraphics[width=12cm]{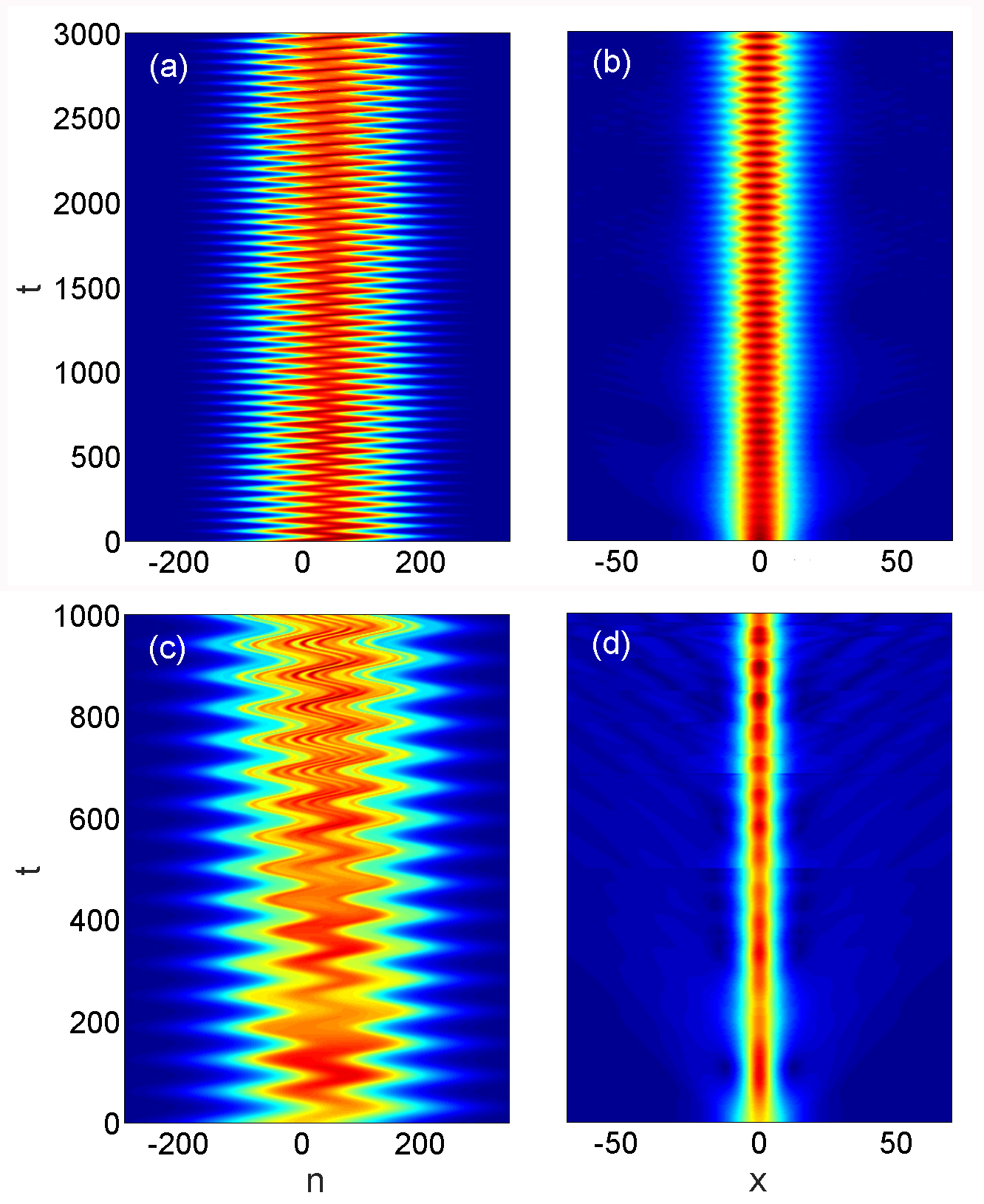}

	\caption{(a) and (b) display the long-time evolution of the wave packet in a nonlinear system in the $n$-$t$ and in the $x$-$t$ planes, respectively.  The system parameters are the same as in Fig.~\ref{fig:one} except for the optimal nonlinearty of $g=0.9$ considered here. (c) and (d) show the destruction of robustness of the BOs in the presence of much stronger nonlinearity $g=1.6$ than the optimal one. The parameters are chosen as $\alpha=0.5$, $\kappa=2$ and $\gamma=0.1$. The initial condition is given by Eq.~(\ref{init}) with $a_0=0.15$ and $w=100$. All the other parameters are identical to those in Fig.~\ref{fig:one} for both cases.}
	\label{fig:three}
\end{figure}

Such robustization is indeed shown in Fig.~\ref{fig:three} (a), (b) which displays the evolution of the wave packet for the same input parameters as shown in Fig.~\ref{fig:one} except that now the nonlinear coefficient of $g=0.9$ is taken into account. Comparing Fig.~\ref{fig:three} (a), (b) to Fig.~\ref{fig:one} (a), (b) clearly demonstrates that the nonlinearity on the one hand prevents the spreading of the wave packet in $x$-direction and on the other hand leads to the existence of long-lived BOs in the $n$-direction. Such evolution can qualitatively be understood to arise from the above explained compensation effect. For our parameters, corresponding to the strongly nonlinear case, the period of the BOs is still very well approximated by the formula  $T=2\pi/\gamma$ derived for the linear case. In particular, for $\gamma=0.1$ the obtained period of the oscillations of the nonlinear wave packet is $\approx 62.8$ which is very close to the oscillations period of the linear case.
The dynamics displayed in Fig.~\ref{fig:three} corresponds to almost $50$ BO periods over which the wave packet is not significantly distorted.

The robustness of the BOs is also confirmed by Fig.~\ref{fig:two}(a) which shows that in the presence of the nonlinearity the deformation parameter $\Delta(t)$ grows with time very slowly. The long-lived BOs require a certain balance of the system parameters to achieve the underlying compensation between diffraction and focusing. The competing effects of the nonlinearity, strength of linear potential and dispersion, are analyzed in Fig.~\ref{fig:two} (b) where we study the spread of the wave packet $\Delta(t)$ after sufficiently long evolution time, more specifically  at $t=3000$, for fixed linear gradients $\gamma$ as function of the strength of the nonlinearity $g$. For each of the studied $\gamma$ we observe clear minima at respective values of the nonlinearity. These minima correspond to the {\em optimal}  relation between the nonlinearity and the linear potential resulting in robust BOs.

In order to get the direct numerical proof of the main result of our paper -- the stabilizing effect of the additional dimension -- we performed the study of the BOs at different values of the diffraction coefficient $\alpha$. Indeed, the limit $\alpha\to 0$ meaning negligible diffraction, returns us to the effectively 1D discrete lattice. Since in this limit the nonlinearity has a destructive effect on BOs, it is natural to expect that the optimal parameter $\Delta(t)$ for smaller $\alpha$ is achieved at smaller nonlinearity $g$, and $g\to 0$ at $\alpha\to 0$. This is exactly what we observe on Fig.~\ref{fig:two}(c). We observe that  increasing $\alpha$ results in an almost linear increase of the optimal $g$, clearly demonstrating that the most robust oscillating regime is achieved when the nonlinearity is balanced by the dispersion. This phenomenon is known to be in the basis of soliton creation in nonlinear systems, and thus allowing us to conjecture that our oscillating object can be viewed as a soliton-like wave packet (see also (\ref{approxim_BO}) and the related discussion).

\begin{figure}
	\includegraphics[width=16cm]{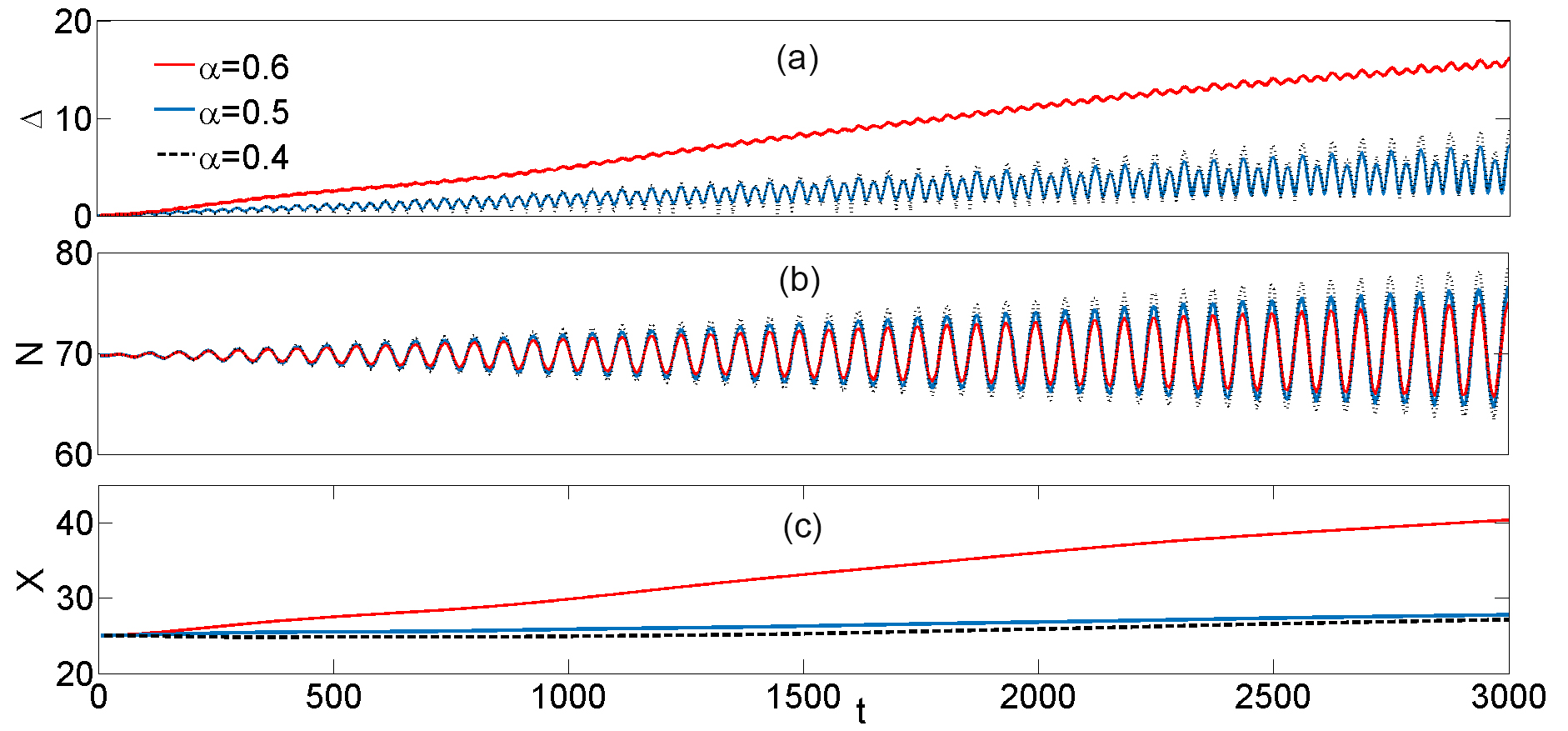}
	\caption{(a) Temporal evolution of the overall spread of the wave packet $\Delta$, (b) N-component of the overall spread of the wave packet $\Delta(t)$, (c) X-component of the overall spread of the wave packet $\Delta$. System parameters are set to $\gamma=0.1$ and $g=0.9$.}
	\label{fig:four}
\end{figure}

Let us now take a closer look at the compromise between the $X$-component and $N$-component of the deformation parameter $\Delta(t)$. To this end we fix  the system parameters optimized for   $\alpha=0.5$ and compare the dynamics of $\Delta(t)$, $N(t)$ and $X(t)$ for these optimal case with the cases where $\alpha=0.4$ and $\alpha=0.6$, i.e. for the evolution at non-optimal diffraction. The results are presented in Fig.~\ref{fig:four}.
  Blue curves show the indicators pertaining to the optimal value of nonlinear coefficient $g=0.9$ for the particular value of $\alpha=0.5$. Two different mechanisms affect each of the components. By increasing the diffraction coefficient $\alpha=0.6$, while maintaining the nonlinear coefficient $g=0.9$, we observe the expected reducing of the wavepacket dispersion $N$ in the "discrete" direction with simultaneous strong increase of the wave packet width $X$ along the continuous directions. Respectively, decreasing the diffraction coefficient $\alpha=0.4$ leads to improvement of the mean square width $X$ with simultaneous increase of $N$. We also note that despite the fact that red and blue curves do not represent optimal cases for the nonlinear regime they still pertain to very robust propagation for a long distance.

\begin{figure}
	\includegraphics[width=16cm]{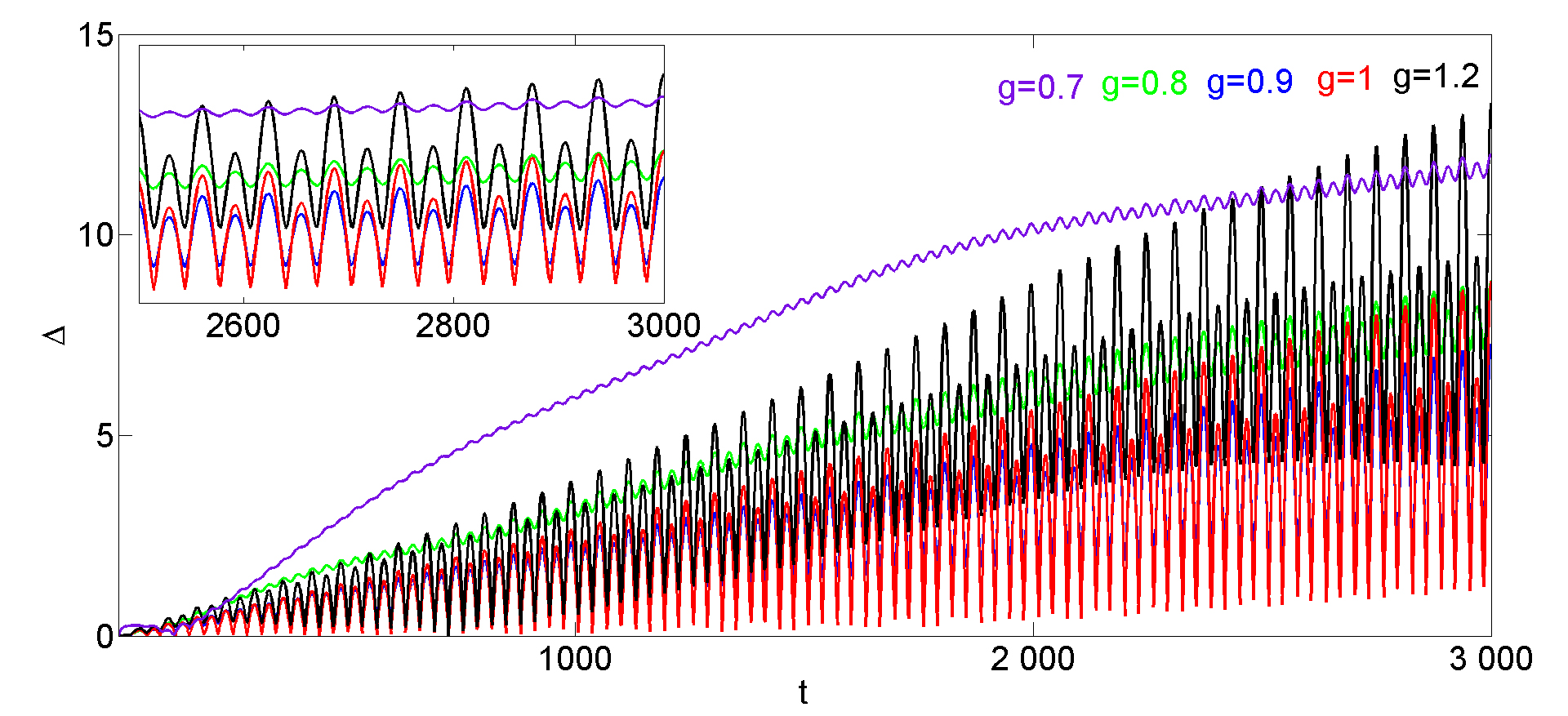}
	\caption{Temporal evolution of the overall spread of the wave packet $\Delta$ for different values of nonlinearity coefficient $g$ with the continuous diffraction coefficient $\alpha=0.5$. The inset shows in details evolution from $t=2500$ to $t=3000$. Gradient strength is set to $\gamma=0.1$.}
	\label{fig:five}
\end{figure}

Figure~\ref{fig:five} shows the uncertainty parameter evolution as a function of the evolution time for five different values of nonlinearity parameter $g$. As we can clearly see all the oscillations are synchronous and this enables us to conclude that results are consistent with evolution to a certain degree. However closer look at the red $g=1$ and blue curves ($g=0.9$ is the optimum) show that for some temporal snapshots the red curve outperform the blue one. This means that strictly speaking we do not have a single point as the optimum but rather some small parameter range where system basically shows optimal behavior. For example the graph presented in Fig. 2(c) would experience very minor fluctuations of its optimal points positions if the integration will be stopped at different time point than $t=3000$. For the sake of experimental realization having broad range of parameters with performance close to optimal is rather advantageous.

We have also performed additional simulations with input different from that defined by Eq.~(3), considering the product of two Gaussians
 \begin{eqnarray}
 \label{init2}
 u_n^{(0)}(x)=a_0e^{-n^2/w^2}e^{-x^2/w_x^2},
 \end{eqnarray}
where $w$ is the characteristic width of the initial wave packet along the $n$-direction, $w_x$ is the characteristic width of the initial wave packet along the $x$-direction, and $a_0$ characterizes the wave packet amplitude. Figure 6 illustrates the evolution with such an input and demonstrates the convergence to a robust propagation regime after initial radiation emission.

\begin{figure}
	\includegraphics[width=12cm]{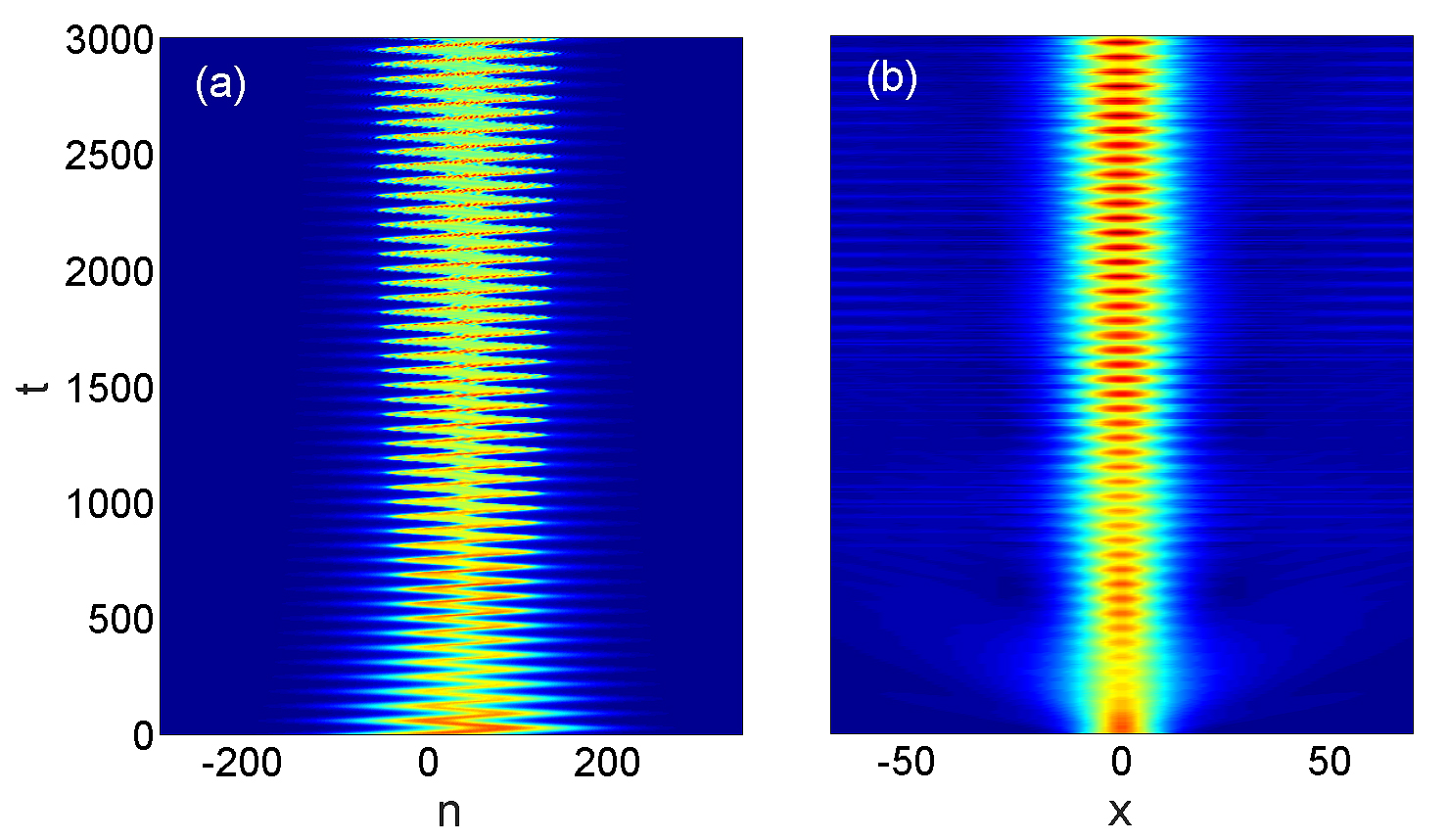}
	\caption{(a) and (b) display the long-time evolution of the wave packet in a nonlinear system in the $n$-$t$ and in the $x$-$t$ planes, respectively.  The system parameters are the same as in Fig.~\ref{fig:one} except for the nonlinearty parameter used is $g=1$.  The initial condition is given by Eq.~(\ref{init2}) with $a_0=0.15$, $w_x=10$ and $w=100$.}
	\label{fig:six}
\end{figure}

Except for the smallest considered gradient of $\gamma=0.01$ we obtain quite broad minima of $\Delta(t=3000)$ as function of $g$
which demonstrates a remarkable robustness of the nonlinear stabilization with respect to change of the nonlinearity.
Returning to Fig.~\ref{fig:three} increase of the nonlinearity above the optimal value, however, may lead to a breakup of the wave packet with a simultaneous compression in the $x$-direction. Such a situation is shown in Fig.~\ref{fig:three} (c), (d) with the nonlinear parameter taken $g=1.6$, that is much higher than $g=0.9$ (the optimal value of the nonlinearity coefficient for the given value of the gradient strength).
\begin{figure}
 	\includegraphics[width=12cm]{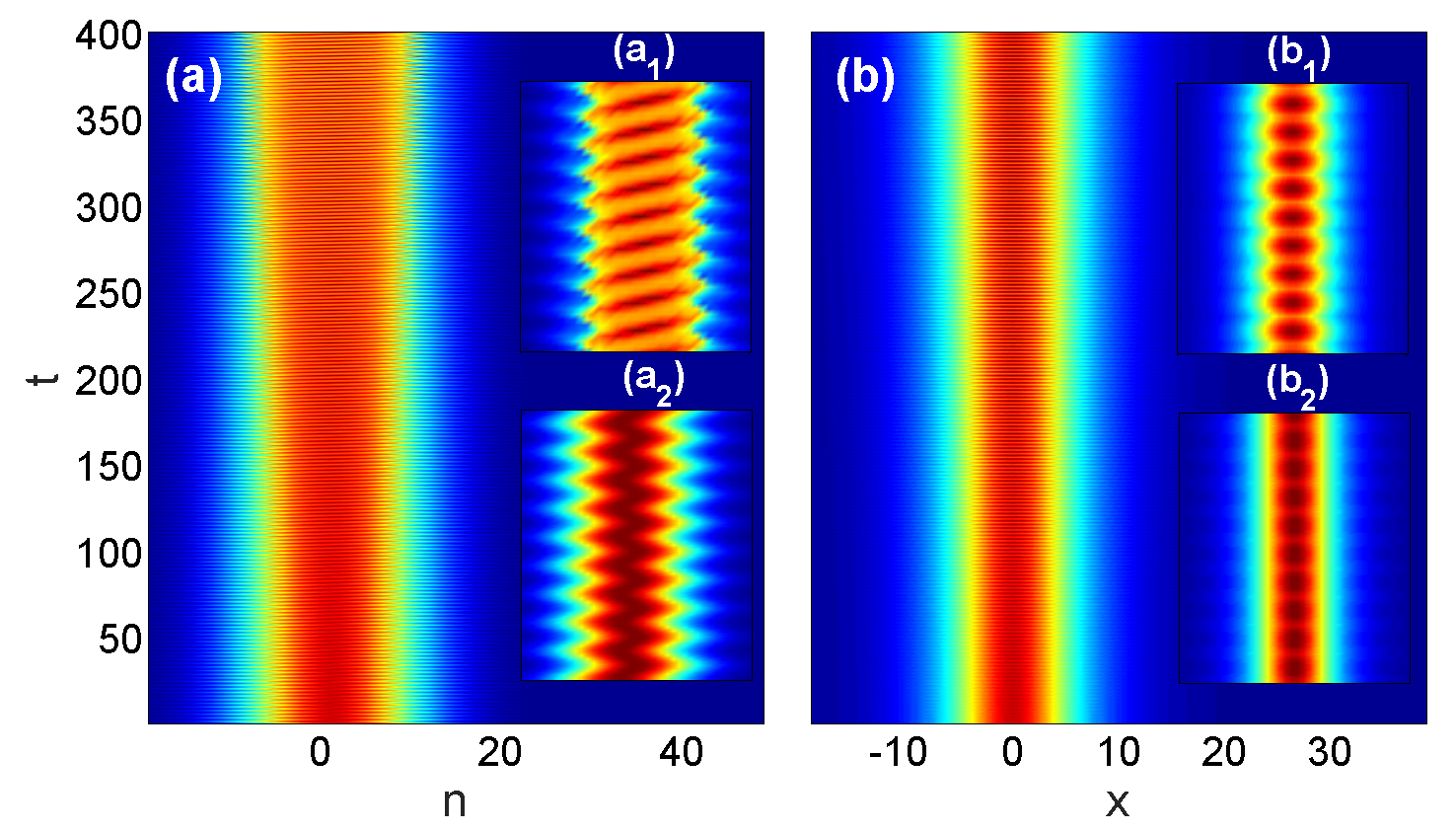}
 	\caption{Panels (a) and (b) show the evolution of the wave packet for a large gradient of $\gamma=3$
in the $n$-$t$ and in the $x$-$t$ planes, respectively.
The shape of the input is as in Eq. (3) and simulation parameters are $\kappa=2$ and $a_0=0.25$.
The insets demonstrate the stable dynamics within few periods from $t=380$ to $t=400$.
The upper insets (a$_1$) and (b$_1$) are obtained from direct numerical solutions of Eq.~(\ref{eq:inf_chain}) and are just zoomed to illustrate the dynamics shown in (a) and (b), respectively.
The lower insets (a$_2$) and (b$_2$) visualize the approximate analytical solution, i.e., Eq.~(\ref{approxim_BO}), shown in the same intervals.}
 	\label{fig:seven}
 \end{figure}

The results reported up to here were obtained for relatively moderate $\gamma$ when the qualitative description could be based on  the band-gap structure of the spectrum which results from the underlying linear lattice.
Meantime, as an alternative view on BOs, a term with the linear gradient strength in a lattice can be transformed in periodically varying coupling coefficients, by a simple gauge transformation, i.e. by the ansatz of $u_n(x,t)\propto\exp{(i\gamma n t})$~\cite{KCV}. Since such a transformation is not directly related to the zone spectrum, it is natural to explore the possibility of obtaining long-lived nonlinear BOs in the case of a relatively large gradient. Fig.~\ref{fig:seven} clearly demonstrates that it is indeed possible to achieve long-lived BOs in the case of a considerable gradient of $\gamma=3$.

As we mentioned above the wave packet with the optimized parameters, whose dynamics is shown in Fig.~\ref{fig:seven} and which manifests remarkable stability, can be viewed as a hybrid of a Bloch oscillating wave and a quasi-soliton. This interpretation is supported by an approximate analytical solution of Eq.~(\ref{eq:inf_chain}). Such approximate solution is obtained by applying the gauge transformation mentioned above for a wave packet that is smooth as function of $n$ allowing to approximate the differences $u_{n\pm1}-u_n$ by their Taylor expansion. It reads
 \begin{eqnarray}
 \label{approxim_BO}
 u_n=\frac{1}{\sqrt{g}} \exp\left[in\gamma t-2i\kappa \left(t+\frac{\sin(\gamma t)}{\gamma}\right) \right]
 \frac{A\left(n+n_0(t)\right)\exp\left[\frac{i}{2}A\left(n+n_0(t)\right) t\right]}{\cosh\left[A\left(n+n_0(t) \right)x\right]},
 \end{eqnarray}
where $n_0(t)=(2{\kappa/{\gamma)}}[1-\cos(\gamma t)]$ defines the location of the center of the wave packet and $A(n)$
describes the wave packet envelope.
Comparing the analytical approximate solution Eq. (6) for the Eq. (1), see Fig.~\ref{fig:seven} (a$_2$) and (b$_2$), with the numerical results, see Fig.~\ref{fig:seven} (a$_1$) and (b$_1$), reveals an excellent overlap for a significant number of oscillation periods. For example, an integral characteristics such as the average width of the oscillations predicted by Eq.~(\ref{approxim_BO}) differs only by about 5\%  up to $t=100$ (corresponding to about 50 oscillation periods)
and the precision drops to a difference of about $40$ percent at $t=400$. The position of the wavepacket is slightly shifted from the center in the course of the evolution in the $n$-plane, see Fig.~\ref{fig:seven}(a$_1$), but remains fixed in the $x$-plane, see Fig.~\ref{fig:seven} (b$_1$). The oscillation period predicted by Eq.~(5) is similar to that obtained from the direct numerical solution of Eq.~(\ref{eq:inf_chain}) as the comparison of the upper and lower insets of Fig.~\ref{fig:seven}(a) demonstrates.

\section*{Conclusions}

To conclude, we have shown that nolinearity is able to support Bloch oscillations when the system is effectively two-dimensional, being discrete, in one dimension and continuous in the orthogonal direction. We have discovered that there exists an optimal relation between nonlinearity and linear gradient strengths allowing for extremely long lived Bloch oscillations (persisting for dozens of oscillation periods with relative deformation of the pulse shape of only a few percents). Such oscillations can be observed even for moderate nonlinearities and large enough values of linear potential, when the band-gap picture of the underlying linear lattice is not applicable anymore. The robust evolution of wave packets in such regime described by an approximate analytical formula in excellent agreement the direct numerical results. The formula describes an object with hybrid features of typical Bloch oscillating wave and soliton.

For future investigations, it would be interesting to analyze a number of points which have not been in the focus of the present study, e.g.,
regarding the interplay between dispersive spreading and decay of the initial pulse in a set of quasi-soliton pulse propagating along the continuous coordinate and the possibilities of observing chaotic regimes,
and achieving asymptotic regimes.

\section*{Acknowledgements}
R.D expresses his gratitude for fruitful discussions to B. A. Malomed. T.M and R.D. acknowledge support of the DFG (Deutsche Forschungsgemeinschaft) through the TRR 142 (project C02) and thank the PC$^2$ (Paderborn Center for Parallel Computing) for providing computing time. The work of VVK  was supported by the FCT (Portugal) grant UID/FIS/00618/2013. A.V.Y acknowledges the support by the Russian Federation Grant 074-U01 through ITMO Early Career Fellowship scheme.

\section*{Author contributions statement}

R.D. and A.Y. initiated the project and performed the simulations. V.V.K derived the analytical results. All the
authors analyzed the data, prepared and reviewed the manuscript. All authors reviewed the manuscript.

\section*{Additional information}
\textbf{Competing financial interests} The authors declare no
competing financial interests.


\end{document}